# AN IMPROVISED FREQUENT PATTERN TREE BASED ASSOCIATION RULE MINING TECHNIQUE WITH MINING FREQUENT ITEM SETS ALGORITHM AND A MODIFIED HEADER TABLE

Vandit Agarwal[1], Mandhani Kushal[2] and Preetham Kumar[3]

[1]Department of Information and Communication Technology, Manipal Institute of Technology, Manipal University, Manipal, Karnataka, India
`vanditagarwal@gmail.com`

[2] Department of Information and Communication Technology, Manipal Institute of Technology, Manipal University, Manipal, Karnataka, India
`kushal.mandhani@gmail.com`

[3] Professor & Head, Department of Information and Communication Technology, Manipal Institute of Technology, Manipal University, Manipal, Karnataka, India
`preetham.kumar@manipal.edu`

## ABSTRACT

*In today's world there is a wide availability of huge amount of data and thus there is a need for turning this data into useful information which is referred to as knowledge. This demand for knowledge discovery process has led to the development of many algorithms used to determine the association rules. One of the major problems faced by these algorithms is generation of candidate sets. The FP-Tree algorithm is one of the most preferred algorithms for association rule mining because it gives association rules without generating candidate sets. But in the process of doing so, it generates many CP-trees which decreases its efficiency. In this research paper, an improvised FP-tree algorithm with a modified header table, along with a spare table and the MFI algorithm for association rule mining is proposed. This algorithm generates frequent item sets without using candidate sets and CP-trees.*

## 1. INTRODUCTION

Data mining [1] is the technique to convert large amounts of data into useful information known as knowledge through a process which can also be described as a knowledge discovery process. It is the result of natural evolution of information technology. The obvious desire for this process of knowledge extraction came from the fact that we are data rich but knowledge poor. Various data mining functionalities are characterization and discrimination, association analysis, cluster analysis, outlier analysis, classification, evolution analysis, prediction etc. Association analysis or association rule mining [2] is one of the most popular data mining techniques. This technique is primarily used to find interesting associations, correlations and similarity in occurrences of various item sets. It helps to reveal many hidden patterns of items in a large dataset. This discovered knowledge helps the managers and top decision makers to make effective decisions. An association rule of the form A $\Rightarrow$ B where A and B are a finite set of items $A_1, A_2, \ldots, A_m$ and $B_1, B_2, \ldots, B_n$ respectively, tells that if in a transaction the set of items A exists, then the set of items B is highly probable to occur in the same transaction. These rules make use of the 2 objective interestingness measures, support and confidence. Support defines the percentage of total transactions which have both the item sets A and B present in one single transaction. Confidence defines the conditional probability that given the presence of item set A in a transaction, with what probability will transaction B be present in the same transaction. An association rule is considered useful only if it meets the minimum threshold defined by support and confidence. Otherwise the association rule is discarded as non-important. Association rule mining finds its major application at every place where the dataset is present or can be converted to a transactional dataset. Its major practical applications include spatial data analysis, share market analysis, XML mining, classification and market basket analysis. This paper presents a novel version of one of the most popular and efficient association mining techniques. The proposed technique makes use of an improvised Frequent Pattern Tree (FP-Tree) and a modified header table along with another table called the spare table and Mining Frequent Item set (MFI) algorithm [3] to efficiently find out the important association rules. The main advantage of the technique involved is that it finds all the frequent item sets in much smaller complexity in terms of both space and time. The outline of this paper is as follows: Section 2 discusses some common contextual notations. Section 3 gives an idea about Improvised FP-Tree. Section 4 throws light on MFI algorithm. Section 5 highlights how to find association rules. Section 6 summarizes the changes and improvements. Sections 7 graphically displays the experimental results obtained on implementing the proposed algorithm.

## 2. CONTEXTUAL NOTATIONS

Even though the data mining system can generate thousands of patterns or rules, not all the patterns or rules are interesting. Interesting here infers that the pattern generated would actually be useful for a decision maker. In this paper two objective measures to find the interestingness of a pattern have been discussed.

### 2.1. Support

This objective measure specifies the probability that a particular transaction contains all the considered items. For instance, there are two items A and B, then their support is the probability that both A and B occur in the same transaction. This can be obtained by finding the ratio of number of transactions containing both A and B and total number of transactions. Thus for a given association rule $A \Rightarrow B$ Support (S) is represented as:

$$S\{A \Rightarrow B\} = \frac{Number\ of\ transactions\ having\ both\ A\ and\ B}{Total\ number\ of\ transactions} \qquad (1)$$

### 2.2. Confidence

This objective measure is used to assess the degree of certainty of the given association rule. This means that a rule $A \Rightarrow B$ holds with confidence x if x% of transactions containing A also contain B.

The confidence of $A \Rightarrow B$ can be calculated by the ratio of number of transactions containing both A and B and total number of transactions containing A. For a given association rule $A \Rightarrow B$, Confidence (C) can be represented as:

$$C\{A \Rightarrow B\} = \frac{No.\ of\ transactions\ having\ both\ A\ and\ B}{Total\ no.\ of\ Transactions\ containing\ A} \qquad (2)$$

In mathematical terms, this can be represented as conditional probability $P\left(\frac{B}{A}\right)$. Consider the following transaction database:

Table 1: Sample Transaction Data.

| Transactions | Items |
|---|---|
| T1 | A,B,D |
| T2 | A,C,D,E |
| T3 | B,D |
| T4 | A |
| T5 | A,B,C,D |

Total number of transactions = 5
Number of transactions having both A and B = 2
Number of transactions having A = 4

$$s\{A \Rightarrow B\} = \frac{2}{5} = 0.4 = 40\%$$

Given a transaction there is a 40% probability that it will contain both A and B or in other words, 40% of the transactions contain both A and B.

$$Confidence\{A \Rightarrow B\} = \frac{2}{4} = 0.5 = 50\%$$

Given a transaction containing A, there is a 50% probability that it will also contain B or in other words, 50% of the transactions containing A also contain B.

These two measures are used as thresholds to determine interestingness of a given association rule is needed.

## 3. IMPROVISED FREQUENT PATTERN TREE

### 3.1. Frequent Pattern (FP) Tree

FP-Tree is a popular method of mining association rules. The entire transactional database is encoded in a compact prefix tree structure. Each transaction is read and represented as a path/branch of the tree. It answers to many shortcomings of the traditional methods of association rule mining like candidate set generation and multiple scans of the database. Each node usually stores three information: name, link and count. Transactions with same item sets or subsets are represented by an overlap in their paths. The parent of all nodes is a "root" node which stores a "NULL" value. This algorithm follows a divide and conquer methodology. A header table structure accompanies the tree in traversal. After the FP-Tree generation, Conditional Pattern (CP) Trees are generated for each item in the FP-Tree to find the frequent item sets.

### 3.2. Key Idea

The main idea behind this paper is to propose an improvised version of the FP-Tree with a modified header table. This FP-Tree is more condensed than the traditional FP-Tree. This reduces the expensive multiple data scans. Also, using the modified header table, we are able to access and check for the existence of any node directly in the tree without having to traverse the entire tree. Instead of CP-Tree, MFI algorithm is used to find the frequent item sets which reduces both time and space complexities.

### 3.3. Modified Header Table

The Modified Header Table consists of count and pointers to the nodes corresponding to each item as we create them. We keep track of the count of each item in the FP-Tree using the header table. Therefore, only the items present in the FP-Tree with their corresponding FP-Tree frequencies are present in the modified header table. So, all the items present in the dataset may not always be present in the modified header table. The items in the header table are present in a descending order of their support.

### 3.4. Proposed Improvised FP-Tree

The proposed FP-Tree algorithm has primarily three data structures - tree, modified header table and spare table [3]. The tree is highly condensed. The tree along with the other two tables represent the entire transactional data set in a highly condensed format. The spare table stores the item name and their count in the table which may be different from the count of the same item in the FP-Tree. Using the modified header table, the need to traverse the tree to search for any element has been eliminated. This reduces the space and time complexities tremendously. Also, since each item has at most one node the FP-Tree, the space required is reduced. The entire process is done in two parts:
  i. FP-Tree construction, spare table generation and modified header table generation.
  ii. MFI algorithm for frequent item set generation.

There are only two cases when an item is added to the spare table [3].
- When the most frequent item is not present in the transaction, all transactions in that transaction are added to the spare table.
- When there is no direct link between the item in consideration and the current root, and the item also exists in the FP-Tree, then it is added to the spare table along with all the following items in that transaction.

### 3.5. Proposed FP-Tree Generation Algorithm

Transaction database is read once to find out the frequency of each item. The most frequent item is identified. The items which do not have the minimum support are removed. All the transactions are arranged in decreasing order of frequency of items in all the transactions. Root node is created and initialized to "NULL". Let the first item of the transaction be $p$ and the remaining items be $q$. If $p$ is the most frequent item, then check if it is the child of root. If not, then create a node corresponding to $p$ as the child of the root node. If it exists then increment the count of the item in the modified header table. Root is shifted to this node. If $p$ is not the most frequent item, then all the items in that transaction are added to spare table. For each item $q$ in the transaction, check if there is a direct link from the current root to this node. If yes, then increment the count in the modified header table and move root to this node. If not, check if the node corresponding to that item exists in the FP tree by looking for that node in the modified header table. If yes, then put the item in spare table. If no, i.e., the node doesn't exist in the FP-Tree, then create a node corresponding to that item as a child of current root. Move the current root to this node. If an item is sent to spare table, then all the following items in that transaction are directly sent to spare table irrespective of any conditions. Repeat until all transactions are read.
The following is the FP-Tree generation algorithm:

**Algorithm 1:** Improvised FP-Tree Generation
**Input**: Transaction database
**Output**: Improvised FP-Tree, modified header table and spare table.
Find the support for each item.
Remove the items which do not meet the minimum support.
Identify the most frequent item in the transaction database.
Create a root node which is referred to as original root.

for each transaction in the database
    Sort the transaction based on the support in a descending order
    Let the first item in each transaction be p and the remaining be q
    Set original root as current root for each transaction.
    if p is the most frequent item
      if p is not child of root
        Create p as the child of current root
        Make the count of the first item in header table as 1
        Make the newly created node as current root
      else
        Make p's node as the current root
        Increase the count of p in the header table
      for all frequent items q when p is the most frequent item
        if q exists as a child of the current node
          Increment the count in the header table
          Move the current root to the child node
        else if q is not present in the header table
          Create a new node for q as the child of current root
          Make the count of the corresponding item in the header table as 1
          Make the newly created node as current root
        else
          Move all remaining items in the transaction to the spare table
    else
      Move all the items of the transaction to the spare table.

### 3.6. Proposed algorithm explanation with example

In order to explain the proposed algorithm, the following example transactional data set is considered.

Table 2: Sample Transaction Data for the FP-Tree

| Transactions | Items |
|---|---|
| T1 | A,B,C |
| T2 | A,D |
| T3 | B,E |
| T4 | A,C |
| T5 | A,C,D |
| T6 | A,B,C,D |
| T7 | A,B,C |

| T8 | B,D |
| T9 | A,B,C,D |

The minimum defined support assumed here is 4. Initially, we scan the transactional dataset and find the support for each item.

Table 3: Transactions Items and their Support

| Item | Support |
| --- | --- |
| A | 7 |
| B | 6 |
| C | 6 |
| D | 5 |
| E | 1 |

Identify the most frequent item. Here A is the most frequent item as its frequency is 7. Identify the items which do not meet the minimum support. Here, E is removed because its frequency is less than the minimum defined support. Create a root node which is referred to as "original root" and initialize it to "NULL". Sorting the transaction based on the support in a descending order is not required in this case as the example transaction database is already sorted. Spare table is initially empty.

1) *Transaction 1: {A,B,C}*

Since {A}, the most frequent item is not the child of original root, we create a new node corresponding to {A} and make it a child of the original root node. Now, {B} is not a child of {A} and {B} is not present in the tree. So a node corresponding to {B} is created and added as a child of node representing {A}. Similarly {C} becomes the child of {B}. Spare table remains empty after this transaction.

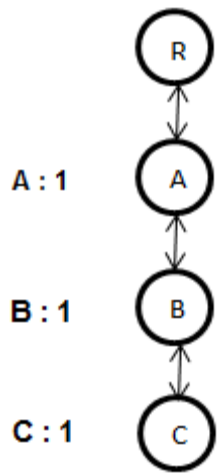

Figure 1: Improvised FP-Tree generation

2) *Transaction 2: {A,D}*

{A} already exists. Therefore, count of {A} is incremented in the modified header table. Since, {D} is not a child of {A} and {D} is not present in the tree, a node corresponding to {D} is created and added as a child of the node representing {A}. Spare table remains empty.

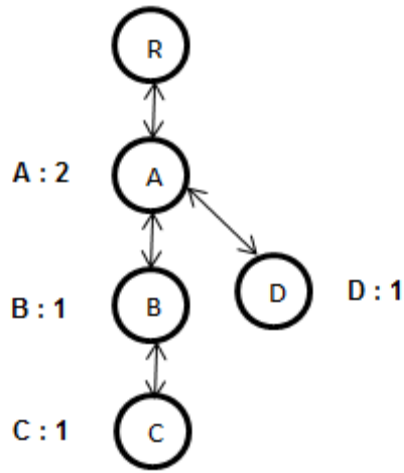

Figure 2: Improvised FP-Tree generation

3) *Transaction 3: {B,E}*

Since {B} is not the most frequent item, all the items in this transaction are sent to the spare table. {E} is not considered because it was removed earlier as it doesn't have the required minimum support.

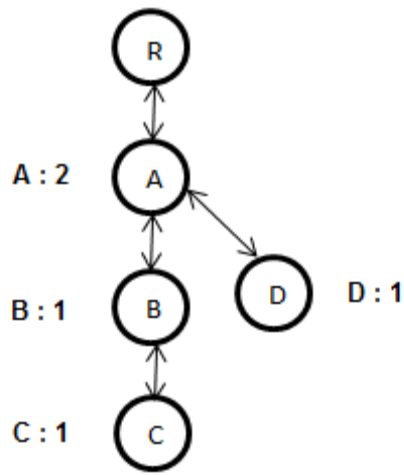

Figure 3: Improvised FP-Tree generation

Table 4: Spare Table

| Item | Frequency |
|------|-----------|
| B    | 1         |

*4) Transaction 4: {A,C}*

{A} already exists. Therefore count of {A} is incremented in the modified header table. Since there is no direct link from {A} to {C} and {C} already exists in the FP-Tree, therefore, {C} is added to the spare table.

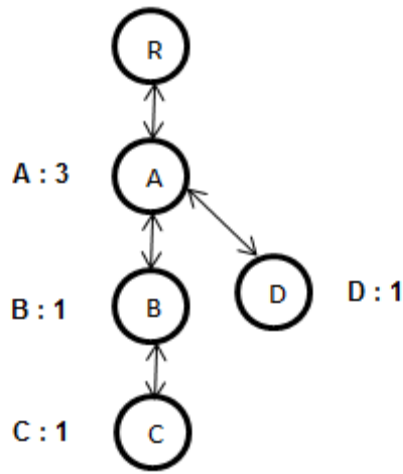

Figure 4: Improvised FP-Tree generation

Table 5: Spare Table

| Item | Frequency |
|------|-----------|
| B    | 1         |
| C    | 1         |

*5) Transaction 5: {A,C,D}*

{A} already exists. Therefore count of {A} is incremented in the modified header table. Since there is no direct link from {A} to {C} as in previous case, all the remaining items including {C} are added to the spare table.

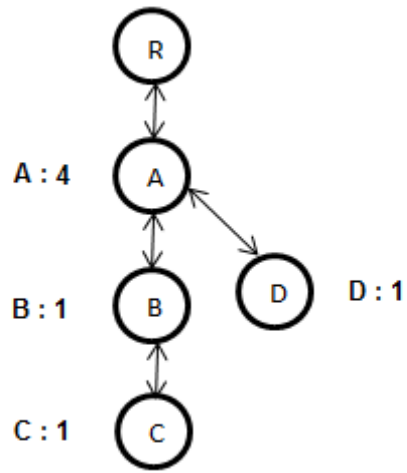

Figure 5: Improvised FP-Tree generation

Table 6: Spare Table

| Item | Frequency |
|------|-----------|
| B    | 1         |
| C    | 2         |
| D    | 1         |

6) *Transaction 9: {A,B,C,D}*

In this case only {D} is added to the spare table. The final FP-Tree and the spare table obtained are shown below.

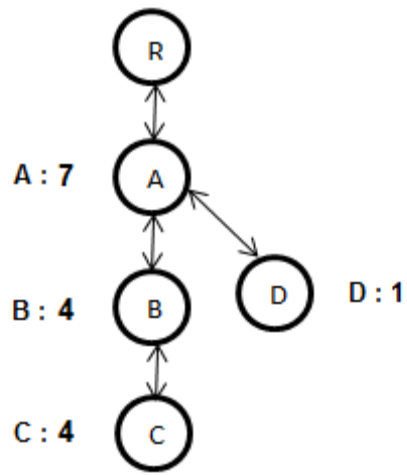

Figure 6: Improvised FP-Tree generation

Table 7: Spare Table

| Item | Frequency |
|------|-----------|
| B    | 2         |
| C    | 2         |
| D    | 4         |

## 4. MINING FREQUENT ITEM SETS (MFI) ALGORITHM

FP-Tree is one of the most efficient and scalable algorithms among the existing association rule mining techniques. This feat is achieved by the FP-Tree algorithm by eliminating the candidate set generation. But in order to do so, it creates a large number of Conditional Pattern (CP) Trees which increases the time and space complexity. The proposed improvised FP-Tree along with the modified header table and the spare table are given as an input to the MFI algorithm [3] which eliminates the necessity to generate the CP-Trees, hence reducing the complexities.

### 4.1. MFI Algorithm Explanation

In this algorithm, the items present in FP-Tree are accessed one by one by traversing the header table since the items in header table are only those present in the FP Tree. This way, it is not required to access the nodes by traversing the tree which needs recursive calls (that in turn increases space and time complexity).

For each item, its frequency is compared with the minimum support. If equal, the item sets with all the elements present at the lower index than the current item's index in the modified header table are formed as the items are placed in descending order in it. If the frequency is less than minimum support, all the items in the FP-Tree which connect this item to the most frequent item (included) are considered. Using these items, the frequent item sets for this item are generated with support as the sum of its frequency and spare table count. If frequency is more, the items are considered in the same manner as above, but with frequency as the frequency of that item in the modified header table. The following is the MFI algorithm.

---

Algorithm 2: Mining Frequent Item sets

Input: FP-Tree, Header table, Stable table

Output: Item set P containing all frequent item sets and their corresponding frequencies.

Initially P is empty.

for all items in the header table

       s = user defined minimum support.

       f = frequency of the item in the header table.

       scount = count in spare table.

       if f is not equal to s

         if f is more than s

              Frequency of frequent item set is FF = f

         else

> Frequency of frequent item set is FF = f + scount
>
> Generate all possible combination of the current item and all the nodes up to most frequent item node in FP-Tree and add them to P with their frequencies as FF
>
> else
>
> Frequency of frequent item set is FF = f
>
> Generate all possible combination of the current item and all the elements present at the lower index than the current item's index in the modified header table.

### 4.2. Experimental Results of MFI Algorithm

Using the example considered in Table 2, the FP-Tree generated in section III, the spare table and header table as the input to the MFI algorithm, the frequent item sets with their frequencies are generated.

The evaluation is started by accessing the items in the modified header table as the items in it are stored in a sorted manner. As explained in the above section, we compare the frequency of the item in the FP-Tree with the minimum support and take action accordingly. In this case, a minimum support of 4 is considered. The first item in the modified header table is {A}. The FP-Tree frequency of this item is 7 which is greater than the minimum support – 4. So the frequency of frequents item sets generated using this item will be the FP-Tree frequency – 7. The frequent items are generated by considering all possible combination of the current item and all the nodes up to most frequent item node in FP-tree. The frequent items generated will be:

$$\{\{A : 7\}\}$$

Now, the second item in the header table is considered – {B} with FP-Tree frequency 4. This means that the FP-Tree frequency and the minimum support are equal. So the frequency of frequents item sets generated using this item will be the FP-Tree frequency – 4. The frequent items in this case are generated by using all possible combinations of the item with all the items in the FP-Tree with a greater or equal frequency. This can be achieved by making all possible combinations of the item with all the items above its index in the modified header table. The modified header table plays a crucial role here as it saves the burden of traversing through the tree to find the items with higher frequency. In this case, there is only one item – {A} present above our current item in the modified header table. So the frequent item sets are:

$$\{\{B : 4\},\{A,B : 4\}\}$$

Similarly, the next item in the FP-Tree also has FP-Tree frequency equal to the minimum support – 4. So the frequency of frequents item sets generated using this item will be the FP-Tree frequency – 4. There are two items – {A},{B} present above the current item in the modified header table. So the frequent item sets are:

$$\{\{C : 4\},\{A,C : 4\},\{B,C : 4\},\{A,B,C : 4\}\}$$

The last item in the modified header is {D} with FP-Tree frequency – 1. The FP-Tree frequency is less than the minimum support. So the frequency of frequent item sets generated using this item will be the FP-Tree frequency (1) added to the spare table count of the item (4), i.e., 1 + 4 = 5. The frequent items are generated using all possible combinations of items up to the most frequent item. The frequent items generated are:

$$\{\{D : 5\},\{A,D :5\}\}$$

The Frequent item sets generated are shown in Table 8.

Table 8: Mining frequent item sets

| Items | Frequent item sets |
|---|---|
| A | {{A : 7}} |
| B | {{B : 4},{A,B : 4}} |
| C | {{C : 4},{A,C : 4},{B,C : 4},{A,B,C : 4}} |
| D | {{D : 5},{A,D :5}} |

## 5. FINDING ASSOCIATION RULES

Considering the minimum confidence threshold as 60%, the following frequent item set is taken to illustrate how the association rules are mined from the frequent item sets.
{{A,B,C : 4}}

1. $A, B \Rightarrow C$
   Frequency {{A,B,C}}= 4
   Frequency {{A,B}}= 4
   Confidence = Frequency {{A,B,C}}/ Frequency {{A,B}}
   Thus, Confidence = 4/4 = 1 =100%. Thus selected.
2. $A, C \Rightarrow B*$
   Frequency {{A,B,C}}= 4
   Frequency {{A,C}}= 4
   Thus, Confidence = 4/4 = 1 =100%. Thus selected.
3. $B, C \Rightarrow A$
   Frequency {{A,B,C}}= 4
   Frequency {{B,C}}= 4
   Thus, Confidence = 4/4 = 1 =100%. Thus selected.
4. $A \Rightarrow B, C$
   Frequency {{A,B,C}}= 4
   Frequency {{A}}= 7
   Thus, Confidence = 4/7 = 0.571 =57.14%. Thus not selected.
5. $B \Rightarrow A, C$
   Frequency {{A,B,C}}= 4
   Frequency {{B}}= 4
   Thus, Confidence = 4/4 = 1 =100%. Thus selected.
6. $C \Rightarrow A, B$
   Frequency {{A,B,C}}= 4
   Frequency {{C}}= 4
   Thus, Confidence = 4/4 = 1 =100%. Thus selected.

In this way, we find the association rules using the frequent item sets generated by MFI algorithm.

## 6. SUMMARIZING THE CHANGES AND IMPROVEMENTS

- Making use of a modified header table. This contains the pointer to each node in the FP tree and respective FP tree frequencies.
- Each item may appear at most once in the FP tree, making it very condensed.

- No tree traversal is required for MFI algorithm. Tree traversal requires recursive calls which increases a lot of space and time complexity. With the discussed implementation, the nodes are accessed in the order in which they are present in the header table, which is essentially just a 1-D array. The items are put in descending order. This reduces the complexity tremendously.
- Number of database scans is less than that in some traditional association rule mining techniques like Apriori algorithm.
- No Conditional Pattern Tree generation.
- No candidate set generation.

## 7. EXPERIMENTAL ANALYSIS

Figure 7 shows the experimental analysis on a data set [4] containing 5665 transactions and 12 items between the traditional FP-Tree [5] and the proposed improvised FP-Tree. The implementation of both the algorithms was done in C using NetBeans IDE 7.4. The data set used in this analysis was obtained from UCI Machine Learning Repository [6].

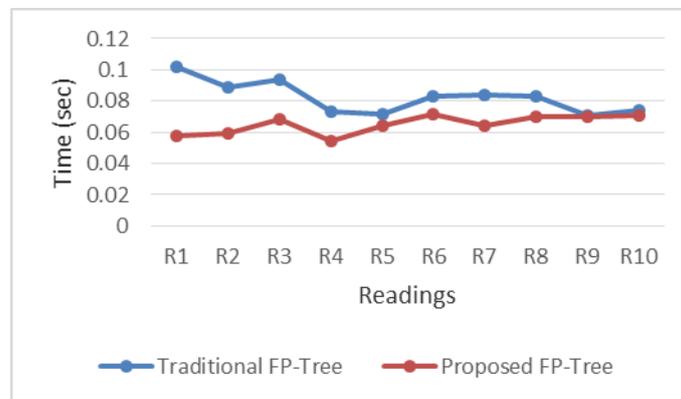

Figure 7: Experimental analysis with 12 items

As can be observed from Figure 7, the proposed improvised FP-Tree algorithm consistently takes less time to find the association rules as compared to the Traditional FP-Tree algorithm. Several readings have been considered in order to analyze the difference in performance between the two algorithms. The minimum support taken is 15%. The mean time taken by the proposed FP-Tree algorithm is 65ms in contrast with 82.3ms taken by the Traditional FP-Tree algorithm.

## 8. CONCLUSIONS

The proposed improvised FP-Tree algorithm contains at maximum one node per item. As a result of this the entire dataset can be represented in a highly condensed manner. Further, by using the modified header table, many recursive calls are avoided in the implementation which reduces both time and space complexities considerably.

**Authors**

| | |
|---|---|
| Vandit Agarwal is a senior undergraduate student pursuing Bachelors of Technology in Information and Communication Technology from Manipal Institute of Technology, Manipal, Karnataka, India. He is enthralled by the limitless existence of possibilities and opportunities that technology has been giving to mankind. His areas of interest include Big Data and Cloud Computing. | 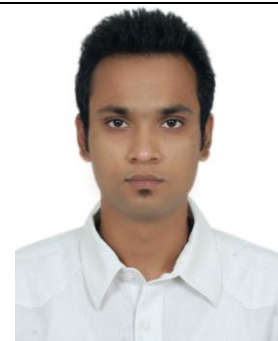 |
| Mandhani Kushal is pursuing Bachelor of Technology degree in Information and Communication Technology from Manipal Institute of Technology, Manipal, India. Exploring new technologies and working with them motivates him. He is currently a final year student in Manipal Institute of Technology, Manipal. His areas of interest are Data Mining and Computer Networks. | 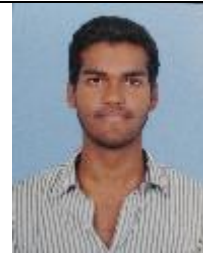 |

Dr. Preetham Kumar is currently employed as the Head of the Department and Professor of Information and Communication Technology at Manipal Institute of Technology, Manipal, Karnataka, India. He has an experience of over 16 years. His areas of interest include Big Data and Operating Systems.

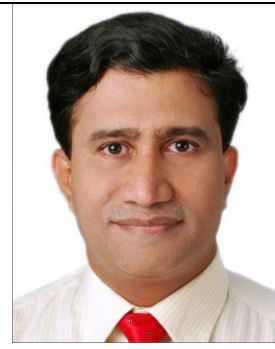